\documentclass[a4paper,fleqn,usenatbib]{mnras}

\usepackage{newtxtext,newtxmath}

\usepackage[T1]{fontenc}
\usepackage{ae,aecompl}

\usepackage{graphicx}

\usepackage{cases}
\usepackage{wasysym}

\newcommand{\rcb}{\mathrm{rcb}}
\newcommand{\hypergeo}{{ }_2 F_1}

\title[Atmospheric mass loss due to giant impacts]{Atmospheric mass loss due to giant impacts: \\the importance of the thermal component for hydrogen-helium envelopes}

\author[J. B. Biersteker and H. E. Schlichting]{John B. Biersteker,$^{1}$\thanks{Email: jo22395@mit.edu}
Hilke E. Schlichting,$^{1,2}$
\\
$^{1}$Massachusetts Institute of Technology, 77 Massachusetts Avenue, Cambridge, MA 02139-4307, USA\\
$^{2}$UCLA, 595 Charles E. Young Drive East, Los Angeles, CA 90095, USA
}

\date{Accepted XXX. Received YYY; in original form ZZZ}

\pubyear{2018}

\begin{document}
\label{firstpage}
\pagerange{\pageref{firstpage}--\pageref{lastpage}}
\maketitle

\begin{abstract}
Systems of close-in super-Earths and mini-Neptunes display striking diversity in planetary bulk density and composition. Giant impacts are expected to play a role in the formation of many of these worlds. Previous works, focused on the mechanical shock caused by a giant impact, have shown that these impacts can eject large fractions of the planetary envelope, offering a partial explanation for the observed spread in exoplanet compositions. Here, we examine the thermal consequences of giant impacts, and show that the atmospheric loss caused by these effects can significantly exceed that caused by mechanical shocks for hydrogen-helium (H/He) envelopes. Specifically, when a giant impact occurs, part of the impact energy is converted into thermal energy, heating the rocky core and the envelope. We find that the ensuing thermal expansion of the envelope can lead to a period of sustained, rapid mass loss through a Parker wind, resulting in the partial or complete erosion of the H/He envelope. The fraction of the envelope mass lost depends on the planet's orbital distance from its host star and its initial thermal state, and hence age. Planets closer to their host stars are more susceptible to thermal atmospheric loss triggered by impacts than ones on wider orbits. Similarly, younger planets, with rocky cores which are still hot and molten from formation, suffer greater atmospheric loss. This is especially interesting because giant impacts are expected to occur $10{-}100~\mathrm{Myr}$ after formation, at a time when super-Earths still retain significant internal heat from formation.
For planets where the thermal energy of the core is much greater than the envelope energy, i.e. super-Earths with H/He envelope mass fractions roughly less than 8 per cent, the impactor mass required for significant atmospheric removal is $M_\mathrm{imp} / M_p \sim \mu / \mu_c \sim 0.1$, approximately the ratio of the heat capacities of the envelope and core. In contrast, when the envelope energy dominates the total energy budget, complete loss can occur when the impactor mass is comparable to the envelope mass.
\end{abstract}


\section{Introduction}
\label{sec: intro}
A major discovery of the Kepler mission is the high abundance of planets intermediate in size between the Earth and Neptune with orbital periods less than 100 days \citep[e.g.][]{2011ApJ...736...19B, 2013ApJ...766...81F, 2013PNAS..11019273P, 2016ApJ...822...86M}. These super-Earths and mini-Neptunes have no obvious solar system analog. Models of exoplanet structure and measurements of the bulk densities of these planets suggest that many possess hydrogen-helium (H/He) envelopes comprising several per cent of the planet's total mass \citep[e.g.][]{2008ApJ...673.1160A, 2014ApJ...783L...6W, 2014ApJ...792....1L, 2015ApJ...801...41R}.

The accretion and evolution of super-Earth H/He envelopes has been studied extensively and can be broadly described by a three step process.
First, planetary cores accrete their H/He envelopes from the protoplanetary nebula \citep[e.g.][]{2015ApJ...811...41L, 2016ApJ...825...29G}; second, the loss of pressure support from the surrounding nebula can lead the initially thermally inflated envelope to shed its outer layers, causing significant mass loss \citep{2016ApJ...817..107O, 2016ApJ...825...29G}; and third, photoevaporation by high energy stellar radiation and the luminosity of the cooling core can erode the atmospheres of close-in planets \citep[e.g.][]{2012ApJ...761...59L, 2013ApJ...775..105O, 2013ApJ...776....2L, 2014ApJ...795...65J, 2018MNRAS.476..759G}.

\begin{figure}
	\centering
	\includegraphics[width=0.5\textwidth]{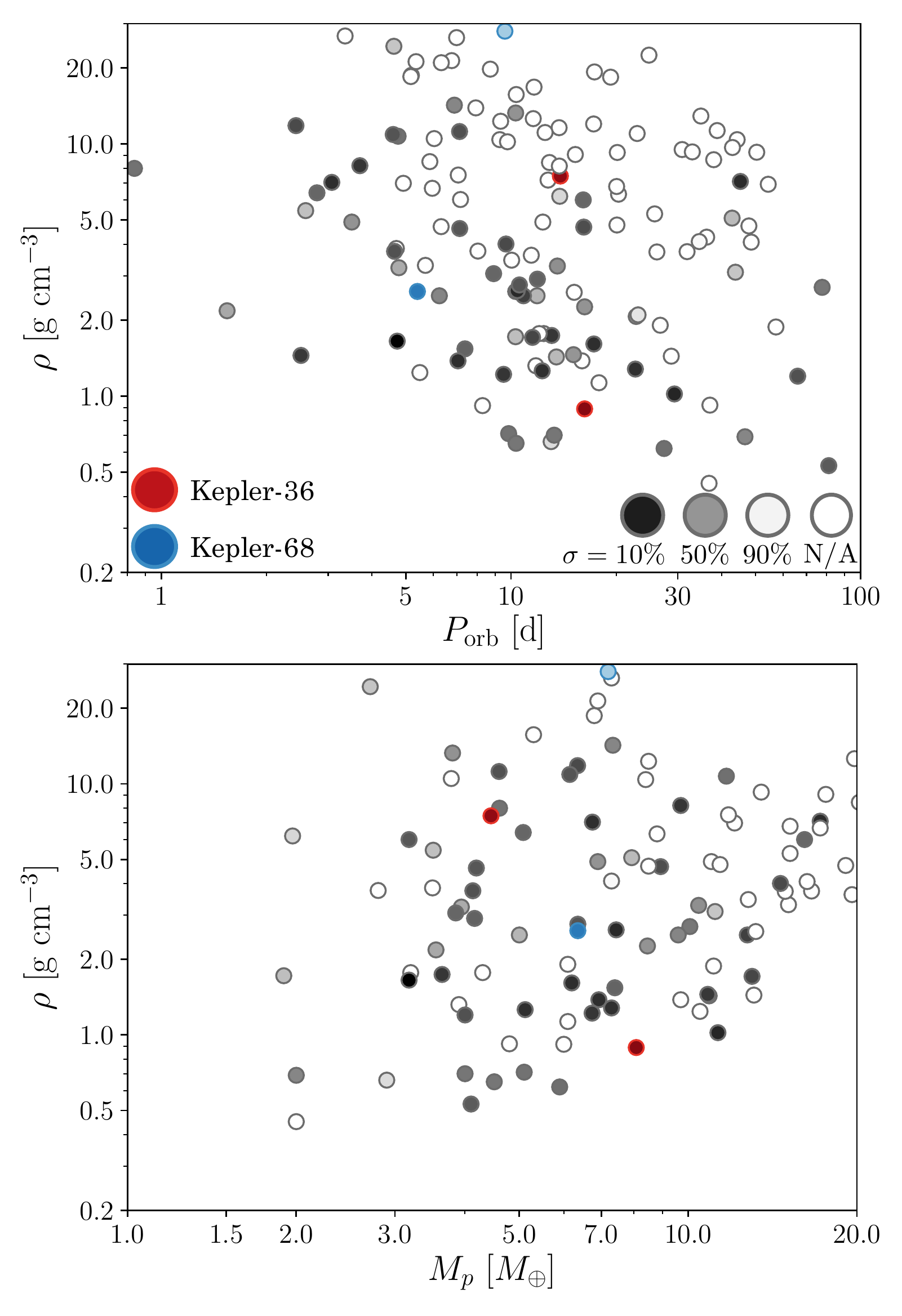}
	\caption{\small Bulk densities ($\rho$) of observed Kepler planets in multi-planet systems with radii ${<}4 R_\oplus$ as a function of orbital period ($P_\mathrm{orb}$) in the top panel, and total planet mass ($M_p$) in the bottom panel, after \citet{2016ApJ...817L..13I}. Red and blue circles highlight Kepler-36 and Kepler-68, respectively, two multi-planet systems with significant density contrast between planets. The transparency indicates the fractional uncertainty in the density, with darker points corresponding to greater certainty. In cases where the planet mass and radius are not provided in the same analysis, the uncertainty on the density is left undefined ($\mathrm{N/A}$) and only an empty circle is plotted. Values are taken from the Composite Kepler planets table at the NASA Exoplanet Archive (June 2018).}
	\label{fig: exoplanet bulk densities}
\end{figure}

Although these processes successfully explain the observed valley in the distribution of small planet radii between $1.5{-}2R_\oplus$ \citep{2017AJ....154..109F, 2018arXiv180501453F, 2017ApJ...847...29O, 2018MNRAS.476..759G}, the exoplanet population displays more compositional diversity than predicted by these processes alone (see Figure \ref{fig: exoplanet bulk densities} and Figure 6 in \citet{2018haex.bookE.141S}), suggesting additional mechanisms are at work. This is particularly the case for planets on orbits outside of ${\sim}45$ days, where photoevaporation and core cooling are less effective, and in multi-planet systems that host planets with dramatically different mean densities \citep{2016ApJ...817L..13I}.

One candidate process for altering planetary bulk densities is giant impacts. Giant impacts are believed to be the last major assembly stage in the formation of the terrestrial planets in our solar system \citep[e.g.][]{2001Icar..152..205C}, and may also play a role in the formation of super-Earths \citep{2013ApJ...775...53H, 2015MNRAS.448.1751I}. Impacts, particularly those which occur after the dissipation of the gas disc, may significantly reduce the H/He envelopes of planets, or strip the cores entirely \citep{2015MNRAS.448.1751I, 2015Icar..247...81S, 2015ApJ...812..164L, 2016ApJ...817L..13I}.

These late giant impacts may be a common occurrence during the formation of super-Earth systems because these planets form in the presence of the gas disc. 
The disc's dynamical interaction with the planets is expected to result in migration and efficient eccentricity damping, producing densely packed planetary systems. Once the gas disc dissipates and its damping effect is removed, these systems may not remain dynamically stable as the planets' eccentricities increase through secular excitation, possibly resulting in orbit crossing and giant impacts before regaining long-term orbital stability \citep{2014A&A...569A..56C, 2017MNRAS.470.1750I, 2018arXiv180200447D}.
Because of the stochastic nature of giant impacts and the small number of impacts expected per system, atmospheric loss through late impacts may offer an especially attractive explanation for the compositional diversity observed among planets in the same system. In this paper, we return to the role of giant impacts in driving volatile loss from super-Earths and mini-Neptunes.

During a giant impact, the impactor is sharply decelerated and its kinetic energy is converted into heat which is deposited at the impact site, creating an event akin to a point-like explosion. This causes a strong shock that propagates through the planet and causes global ground motion, launching a shock into the planet's envelope. A series of previous studies of giant impacts have focused on the ejection of the atmosphere by this hydrodynamic shock \citep{2003Icar..164..149G, 2015Icar..247...81S, 2016ApJ...817L..13I}. In addition, \citet{2015ApJ...812..164L} investigated volatile loss in two specific giant impact scenarios using 3D numerical simulation. They find that giant impacts can lead to significant loss of volatiles and point out that the atmospheric loss in their simulation is likely further enhanced by Parker winds and photoevaporation of the thermally inflated post-impact envelope.

In the case of H/He envelopes, the heating of the core during an impact can lead to significant mass loss. Heat from the core is transferred to the envelope, either directly unbinding it, or leading to significant thermal expansion and gradual mass loss via Parker winds.
In this paper, we focus specifically on the thermal aspect of giant impacts, examining the energy of the impact, the thermal inflation of the H/He envelope after the impact, and the subsequent atmospheric loss. We demonstrate that the thermal component of the atmospheric loss can dominate the loss caused by shocks for H/He atmospheres and that it may partly explain the observed compositional diversity in super-Earth systems. This paper is structured as follows: we begin with a model for the structure and evolution of a super-Earth envelope and core in Section \ref{sec: atm struct and evolution}, calculate the atmospheric mass loss in a range of impact scenarios in Section \ref{sec: impact mass loss}, and conclude with a discussion in Section \ref{sec: discussion}.

\section{Envelope structure and evolution}
\label{sec: atm struct and evolution}
\begin{figure*}
	\centering
	\includegraphics[width=\textwidth]{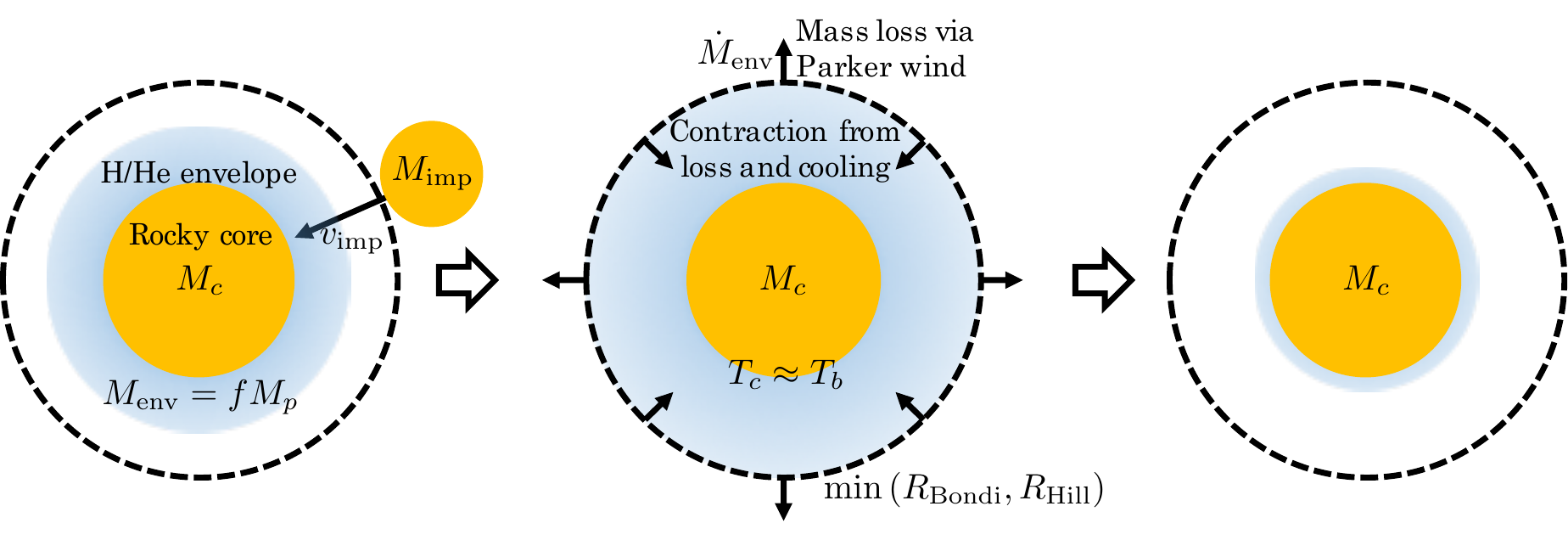}
	\caption{\small Schematic of atmospheric mass loss following a giant impact. In the left panel, an impactor with mass $M_\mathrm{imp}$ strikes a planet of mass $M_p$ at impact velocity $v_\mathrm{imp}$. The planet has a core mass of $M_c \approx M_p$ and envelope mass $M_\mathrm{env} = f M_p$. In the middle panel, heat from the impact has inflated the envelope so that it loses mass through a Parker wind at its outer radius $R_\mathrm{out} = \min{(R_B, R_H)}$. Mass loss and cooling cause the envelope to contract, eventually halting appreciable atmospheric loss. At right, the planet is left to continue cooling with a significantly reduced envelope.}
	\label{fig: cartoon explanation}
\end{figure*}

Giant impacts deposit significant energy into a planet's core and envelope, leaving the envelope in a thermally inflated state which can be highly susceptible to hydrodynamic escape (Figure \ref{fig: cartoon explanation}).
The exact rate of mass loss from the planetary envelope is determined by the atmospheric density at the outer edge of the envelope, where gas molecules are only tenuously gravitationally bound to the planet. This outer radius, $R_\mathrm{out}$, is the lesser of the Bondi radius, $R_B = 2 G M_p / c_s^2$, and the Hill radius, $R_H = a \left( M_p / (3 M_*) \right)^{1/3}$, where $G$ is the gravitational constant, $M_p$ is the planet mass, $M_*$ is the stellar mass, $a$ is the semimajor axis of the planet's orbit, and $c_s$ is the sound speed of the gas. To determine the density at this radius, we first develop a simple model of the planet's envelope and core appropriate for conditions following a large impact. We then consider the evolution of this structure as the planet sheds mass, cools, and contracts to determine the total mass lost from the envelope.

\subsection{Envelope structure model}
\label{sec: atm struct}
We model the planet's envelope as an inner convective region with an adiabatic profile, and an outer radiative layer with an isothermal profile \citep[see][]{2006ApJ...648..666R,2014ApJ...786...21P,2015MNRAS.448.1751I,2016ApJ...825...29G}. In all the cases we consider here, the mass of the planet is dominated by the mass of the rocky core, so that $M_p \sim M_c$. Neglecting, therefore, the self-gravity of the envelope, the density, pressure, and temperature in the convective portion of the envelope are given by
\begin{align}
\frac{\rho}{\rho_b} & = \left[ \frac{\gamma - 1}{\gamma} \Lambda \left( \frac{R_c}{r} - 1 \right) + 1 \right]^{\frac{1}{\gamma - 1}} \text{,}
\label{eq: density}
\\
\frac{P}{P_b} & = \left[ \frac{\gamma - 1}{\gamma} \Lambda \left( \frac{R_c}{r} - 1 \right) + 1 \right]^{\frac{\gamma}{\gamma - 1}} \text{,}
\\
\frac{T}{T_b} & = \frac{\gamma - 1}{\gamma} \Lambda \left( \frac{R_c}{r} - 1 \right) + 1 \text{,}
\end{align}
where $R_c$ is the core radius, $r$ is the distance from the centre of the core, and $\gamma$ is the adiabatic index of the envelope. We define $\Lambda$ as approximately the ratio of the gravitational potential and kinetic energies of a gas particle at the base of the envelope, $\Lambda \equiv G M_c \mu / (R_c k_B T_b)$, where $\mu$ is the mean molecular weight of the envelope and $k_B$ is the Boltzmann constant. The density, pressure, and temperature at the base of the envelope are $\rho_b$, $P_b$, and $T_b$, respectively.

At large enough radii, the atmospheric density decreases to the point that the envelope becomes radiative. In this region, the envelope is nearly isothermal with a temperature which is approximately the equilibrium temperature:
\begin{align}
T_\mathrm{eq} = \left( \frac{1 - A_B}{4} \right)^{1/4} \sqrt{\frac{R_{*}}{a}} T_{*} \text{,}
\end{align}
where $A_B$ is the Bond albedo, $R_*$ is the stellar radius, and $T_*$ is the star's effective temperature. Assuming the envelope outside this radiative-convective boundary (RCB) to be completely isothermal, the density profile of the envelope becomes exponential,
\begin{align}
\label{eq: isothermal density}
\rho & = \rho_\mathrm{rcb} \exp{\left[ \frac{R_\rcb}{h} \left(\frac{R_\rcb}{r} - 1 \right) \right]} \text{,}
\end{align}
with a scale height $h = k_B T_\mathrm{eq} R_\rcb^2 / (G M_p \mu)$, where $r = R_\rcb$ is the radius of the radiative-convective boundary and $\rho_\rcb = \rho{(R_\rcb)}$. The pressure is then, using the ideal gas law, $P = \rho k_B T_\mathrm{eq} / \mu$.

The mass in the convective region of the envelope is given by integrating Equation \eqref{eq: density}:
\begin{align}
M_\mathrm{env} & = \int_{R_c}^{R_\rcb} 4 \pi r^2 \rho_b \left[\frac{\gamma - 1}{\gamma} \Lambda \left( \frac{R_c}{r} - 1 \right) + 1 \right]^{\frac{1}{\gamma - 1}} {dr}
\\
& = 4 \pi \rho_b \frac{\gamma - 1}{4 - 3\gamma} R_c^3 ( \nabla_{\mathrm{ad}} \Lambda )^{\frac{1}{\gamma - 1}} \nonumber
\\
& \qquad \times \left[ \hypergeo \left( \theta_m; \frac{\nabla_\mathrm{ad} \Lambda - 1}{\nabla_\mathrm{ad} \Lambda}  \right) \right. \nonumber
\\
& \qquad \qquad - \left. \left( \frac{R_c}{R_\rcb} \right)^\frac{4 - 3 \gamma}{\gamma - 1} \hypergeo \left( \theta_m; \frac{\nabla_\mathrm{ad} \Lambda - 1}{\nabla_\mathrm{ad} \Lambda} \frac{R_\rcb}{R_c} \right) \right]
\label{eq: env mass}
\end{align}
where ${ }_2 F_1 $ is the ordinary hypergeometric function, 
$\theta_m = \left[ (3\gamma - 4)/(\gamma - 1), -1/(\gamma - 1); (4\gamma - 5)/(\gamma - 1) \right]$, and $\nabla_\mathrm{ad} \equiv (\gamma - 1)/\gamma$. The energy of the envelope is determined from the gravitational and thermal energies, $E_\mathrm{env} = E_g + E_\mathrm{th}$, where the gravitational energy for a non-self-gravitating envelope is
\begin{align}
E_g & = -4 \pi G M_c \rho_b \int_{R_c}^{R_\rcb} \left[\nabla_\mathrm{ad} \Lambda \left( \frac{R_c}{r} - 1 \right) + 1 \right]^{\frac{1}{\gamma - 1}} r {dr}
\\
& = -4 \pi G M_c \rho_b R_c^2 \left( \frac{\gamma - 1}{3 - 2 \gamma} \right) (\nabla_{\mathrm{ad}} \Lambda )^{\frac{1}{\gamma - 1}} \nonumber
\\
& \qquad \times \left[ \hypergeo \left( \theta_g; \frac{\nabla_\mathrm{ad} \Lambda - 1}{\nabla_\mathrm{ad} \Lambda} \right) \right. \nonumber
\\
& \qquad \qquad \left. - \left( \frac{R_c}{R_\rcb} \right)^{\frac{3 - 2 \gamma}{\gamma - 1}} \hypergeo \left( \theta_g; \frac{\nabla_\mathrm{ad} \Lambda - 1}{\nabla_\mathrm{ad} \Lambda} \frac{R_\rcb}{R_c} \right) \right] \text{,}
\label{eq: grav energy}
\end{align}
where 
$\theta_g = \left[ (2 \gamma - 3) / (\gamma - 1), -1 / (\gamma - 1); \right.$ $\left. (3 \gamma - 4)/(\gamma - 1) \right]$.
 And the thermal energy is
\begin{align}
E_\mathrm{th} & = \frac{4 \pi \rho_b T_b k_B}{\mu (\gamma - 1)} \int_{R_c}^{R_\rcb} \left[ \frac{\gamma - 1}{\gamma} \Lambda \left( \frac{R_c}{r} - 1 \right) + 1 \right]^{\frac{\gamma}{\gamma - 1}} r^2 {dr}
\\
& = \frac{4 \pi \rho_b T_b k_B R_c^3}{\mu (\gamma - 1)} \left( \frac{\gamma - 1}{3 - 2\gamma} \right) (\nabla_{\mathrm{ad}} \Lambda )^{\frac{\gamma}{\gamma - 1}} \nonumber
\\
& \qquad \times \left[ \hypergeo \left(\theta_\mathrm{th}; \frac{\nabla_{\mathrm{ad}} \Lambda - 1}{\nabla_{\mathrm{ad}} \Lambda} \right) \right. \nonumber
\\
& \qquad \qquad \left. - \left( \frac{R_c}{R_\rcb} \right)^{\frac{3 - 2\gamma}{\gamma - 1}} \hypergeo \left(  \theta_\mathrm{th}; \frac{\nabla_{\mathrm{ad}} \Lambda - 1}{\nabla_{\mathrm{ad}} \Lambda} \frac{R_\rcb}{R_c} \right) \right] \text{,}
\label{eq: thermal energy}
\end{align}
where 
$\theta_\mathrm{th} = \left[ (2 \gamma - 3) / (\gamma - 1), -\gamma / (\gamma - 1); \right.$ $\left. (3 \gamma - 4) / (\gamma - 1) \right]$. 
In cases considered in this paper, when $T_b$ significantly exceeds $T_\mathrm{eq}$, most of the envelope's mass is in the inner convective region. Ignoring the mass in the isothermal layer, the atmospheric structure is then entirely determined by the envelope mass (Equation \ref{eq: env mass}) and the envelope energy.

\subsubsection{Core model}
The planets we are considering are young ($10{-}100~\mathrm{Myr}$) and have significant H/He envelopes. This allows them to retain significant heat from their formation so that the basal temperature of the envelope is ${\gtrsim}2000~\mathrm{K}$, higher than the melting point of silicates. We therefore assume the core is undifferentiated, fully molten with no solid insulating crust, and that heat is efficiently transferred between the core and the base of the envelope so that the base temperature and core temperature are the same ($T_b \approx T_c$).

Adiabatic profiles of terrestrial planet interiors show changes in temperature of a factor of only a few over changes in depth of thousands of kilometers \citep{2010PEPI..183..212K}, so we approximate the core as isothermal with an energy
\begin{align}
E_c  \sim c_{V,c} M_c T_c \text{,}
\label{eq: core energy}
\end{align}
where $c_{V,c} \sim 5 - 10 \times 10^2~\mathrm{J}~\mathrm{kg}^{-1}~\mathrm{K}^{-1}$ is the specific heat of the core \citep{2001PhRvB..64d5123A, 1995ApJ...450..463G,2012ApJ...761...59L}. Finally, we adopt $R_c / R_\oplus = (M_c / M_\oplus)^{1/4}$ as the core mass-radius relationship \citep[e.g.][]{2006Icar..181..545V}.

\subsection{Envelope evolution}
\label{sec: envelope evolution}
After an impact the envelope is inflated, enabling enhanced mass loss. As the envelope cools and contracts, the density at the outer radius ($R_\mathrm{out}$) drops, quenching the mass loss. The thermal evolution of the envelope is driven by both radiation and the loss of energy through mass loss \citep{2016ApJ...817..107O, 2016ApJ...825...29G, 2018MNRAS.476..759G}. The mass loss rate is approximately
\begin{align}
\dot{M}_\mathrm{env} \approx -4 \pi R^2_\mathrm{out} \rho_\mathrm{out} c_s \text{,}
\label{eq: mdot}
\end{align}
where $c_s = \sqrt{\gamma k_B T / \mu}$ is the sound speed. Raising a gas parcel from the outer envelope to infinity requires a corresponding change in the envelope's energy,
\begin{align}
\dot{E}_\mathrm{env,m} \approx \frac{G M_c \dot{M}_\mathrm{env}}{R_\mathrm{out}} \text{.}
\end{align}
The planet also cools radiatively, with an approximate luminosity given by combining the equations of flux conservation and hydrodynamic equilibrium at the RCB:
\begin{align}
\dot{E}_\mathrm{env,L} = -L_\rcb = -\nabla_\mathrm{ad} \frac{64 \pi \sigma T_\rcb^3 G M_c \mu}{3 \kappa_R \rho_\rcb k_B} \text{,}
\end{align}
where $\sigma$ is the Stefan-Boltzmann constant and $\kappa_R$ is the Rosseland mean opacity of the envelope at the RCB. We adopt a value of $\kappa_R = 0.1~\mathrm{cm}^2~\mathrm{g}^{-1}$ which is an appropriate approximation for the temperatures and conditions we expect at the RCB \citep{2008ApJS..174..504F}.

\begin{figure}
	\centering
	\includegraphics[width=0.5\textwidth]{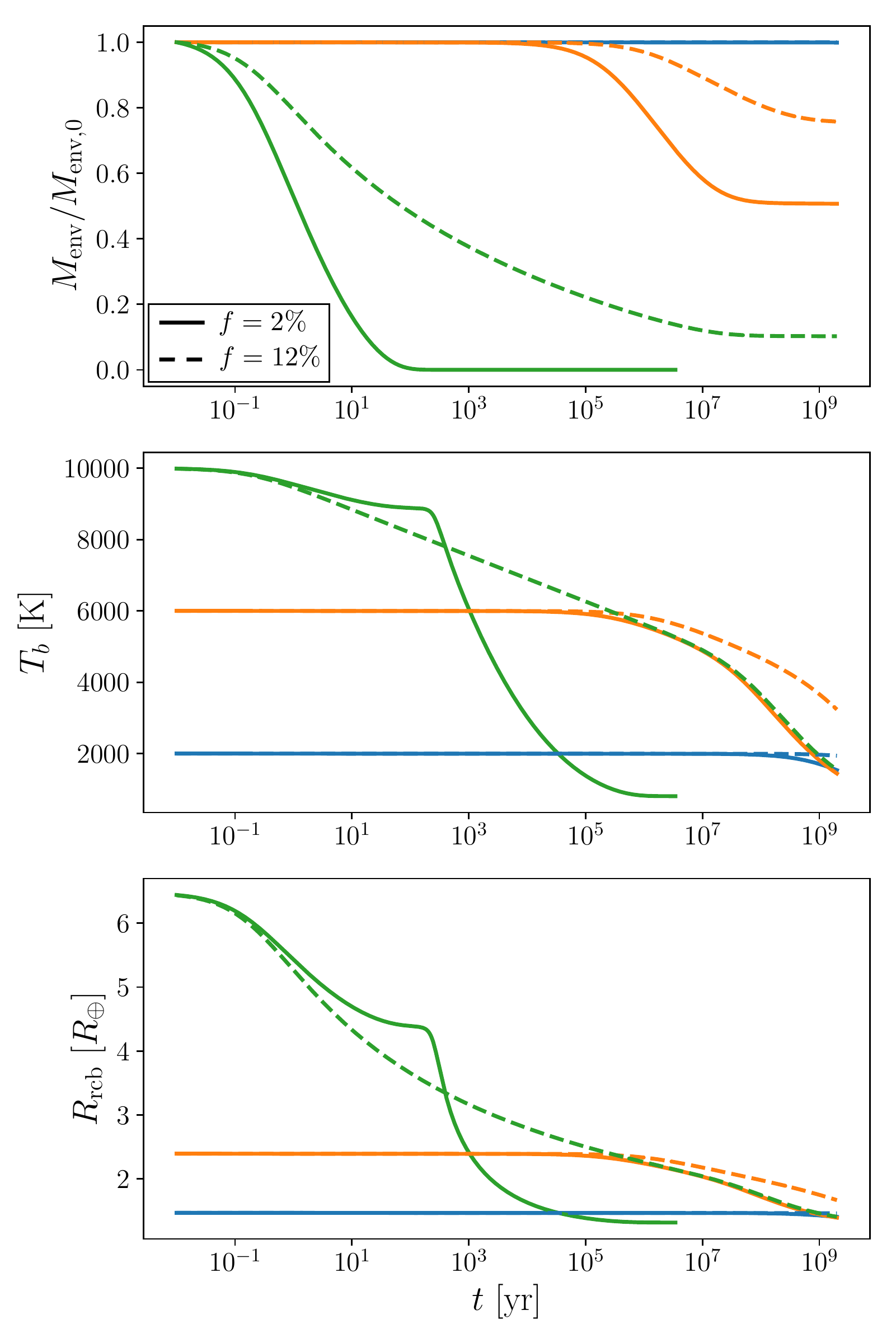}
	\caption{\small Examples of the evolution of the mass, base temperature, and radiative-convective boundary location of a H/He envelope for a planet with $M_c = 3 M_\oplus$, $R_c = 1.32 R_\oplus$, $a = 0.1~\mathrm{au}$, $f = 2$ per cent, and a range of initial base temperatures, $T_{b,0}$. The green, orange, and blue lines correspond, respectively, to initial base temperatures of $2000$, $6000$, and $10000~\mathrm{K}$. The solid and dashed lines correspond to envelope mass fractions of $f=2$ and $12$ per cent.}
	\label{fig: example envelope evolution}
\end{figure}

Combining the equations describing the envelope and core energies (Equations \ref{eq: grav energy}, \ref{eq: thermal energy}, and \ref{eq: core energy}) yields the total energy $E = E_g + E_\mathrm{th} + E_c$, which evolves by $\dot{E} = \dot{E}_\mathrm{env, m} +\dot{E}_\mathrm{env, L}$. The core mass is fixed and the envelope mass diminishes according to Equation \eqref{eq: mdot}. As described in Section \ref{sec: atm struct}, the envelope structure is entirely determined by its mass and the energy of the system, so these equations are sufficient to describe the combined thermal and mass loss evolution of the planet (Figure \ref{fig: example envelope evolution}).

\section{Mass Loss From Giant Impacts}
\label{sec: impact mass loss}
Giant impacts during late planetary accretion can significantly alter the atmosphere of a nascent super-Earth. A large impact generates a shockwave which propagates through the planet, causing global ground motion which can immediately eject a fraction of the envelope \citep[e.g.][]{2015Icar..247...81S}. Additionally, the impactor delivers significant energy to the planet, heating the core, which in turn heats the envelope. The thermally inflated envelope can then be vulnerable to hydrodynamic escape and lose mass rapidly.
Focusing on the thermal component of the mass loss, we derive analytical estimates for both the time-scale of atmospheric loss and the impactor mass required to remove the atmosphere entirely. We then verify these estimates and explore a range of parameters using the numerical model described in Section \ref{sec: atm struct and evolution}.

\subsection{Analytical estimates}
If an impactor delivers enough energy, a planet's gaseous envelope can be entirely removed. The energy required depends on the relative shares of the envelope and core in the energy budget of the planet. Using Equation \eqref{eq: core energy} for the core's energy and taking $E_\mathrm{env} \sim N k_B T_c / (\gamma - 1)$ as the approximate thermal energy of the envelope, the envelope mass fraction marking the transition between the core-dominated and envelope-dominated regimes for a H/He envelope is
\begin{align}
 f_b = \left( \frac{k_B}{(\gamma - 1) \mu c_{V,c}} + 1 \right)^{-1} \approx 0.08 \text{,}
\end{align}
where $\mu = 2.34~\mathrm{u}$ is the mean molecular weight of the envelope, $\gamma = 7/5$, and $c_{V,c} = 750~\mathrm{J}~\mathrm{K}^{-1}~\mathrm{kg}^{-1}$.

In the limit where the core energy dominates ($f \ll f_b$), the envelope will be entirely lost when the temperature at the base of the envelope is so high that gas molecules do not remain gravitationally bound; equivalently, when the Bondi radius equals the core radius. Using the previously adopted core mass-radius relationship, $R_c / R_\oplus = (M_c / M_\oplus)^{1/4}$, the required core temperature for complete loss is
\begin{align}
\label{eq: core temperature}
T_c & = \frac{2 G M_c \mu}{\gamma k_B R_c}
\\
& \approx 25,000~\mathrm{K} \times \left( \frac{M_c}{M_\oplus} \right)^{3/4} \left( \frac{\mu}{2.34~\mathrm{u}} \right) \left( \frac{7/5}{\gamma} \right)\text{.}\nonumber
\end{align}
The energy delivered by the impactor is $E_\mathrm{imp} = \eta M_\mathrm{imp} v_\mathrm{imp}^2 / 2$, where $\eta \in [0, 1]$ is the fraction of the total impact energy available for heating the core and envelope. Assuming an impact velocity $v_\mathrm{imp} \sim v_\mathrm{esc}$, the energy is then
\begin{align}
\label{eq: impact energy}
E_\mathrm{imp} \sim \eta M_\mathrm{imp} \frac{G M_p}{R_c} \text{.}
\end{align}
The corresponding impactor mass, neglecting the initial temperature of the core, is 
\begin{align}
\label{eq: impactor mass core dominated}
M_\mathrm{imp} & \sim \frac{2 \mu c_{V,c}}{\eta \gamma k_B} M_c
\\
& \sim \frac{2}{\eta \gamma (\gamma_c - 1)} \frac{\mu}{\mu_c} M_c \nonumber
\end{align}
where, in the second expression, $\mu_c$ and $\gamma_c$ are the core's mean molecular weight and adiabatic index. 
For core-dominated energy budgets, the impactor mass required for atmospheric loss is approximately determined by the envelope-to-core mean molecular weight ratio, making H/He envelopes particularly susceptible to removal.
For an impact efficiency $\eta~\sim~1$, total atmospheric loss is predicted by Equation \eqref{eq: impactor mass core dominated} at $M_\mathrm{imp} \sim 0.3 M_c$.

When the starting temperature of the core is comparable to the ejection temperature, then the impactor only needs to deliver enough energy to raise the core temperature $\Delta{T} = T_c - T_0$. The corrected impactor mass is then given by multiplying the right-hand side of Equation \eqref{eq: impactor mass core dominated} by $\Delta{T} / T_c$. In the case of recently formed super-Earths which still retain significant heat from formation, $\Delta{T}$ may be small, so that complete atmospheric loss may occur for lower impactor masses.

When the core's share of the energy budget can be neglected to first order ($f \gg f_b$), then the envelope can be ejected when $E_\mathrm{imp} + E_\mathrm{env} = 0$. Approximating the envelope energy as $-G M_\mathrm{env} M_c / R_c$, an impactor mass of
\begin{align}
\label{eq: env eject mass}
M_\mathrm{imp} \sim \frac{1-f}{\eta} M_\mathrm{env} = \frac{f}{\eta} M_c
\end{align}
is required to eject the whole atmosphere. Equivalently, for an efficiency $\eta \sim 1$, the impactor mass must be comparable to the mass of the envelope.

\subsubsection{Partial atmospheric loss}
\begin{figure}
	\centering
	\includegraphics[width=0.5\textwidth]{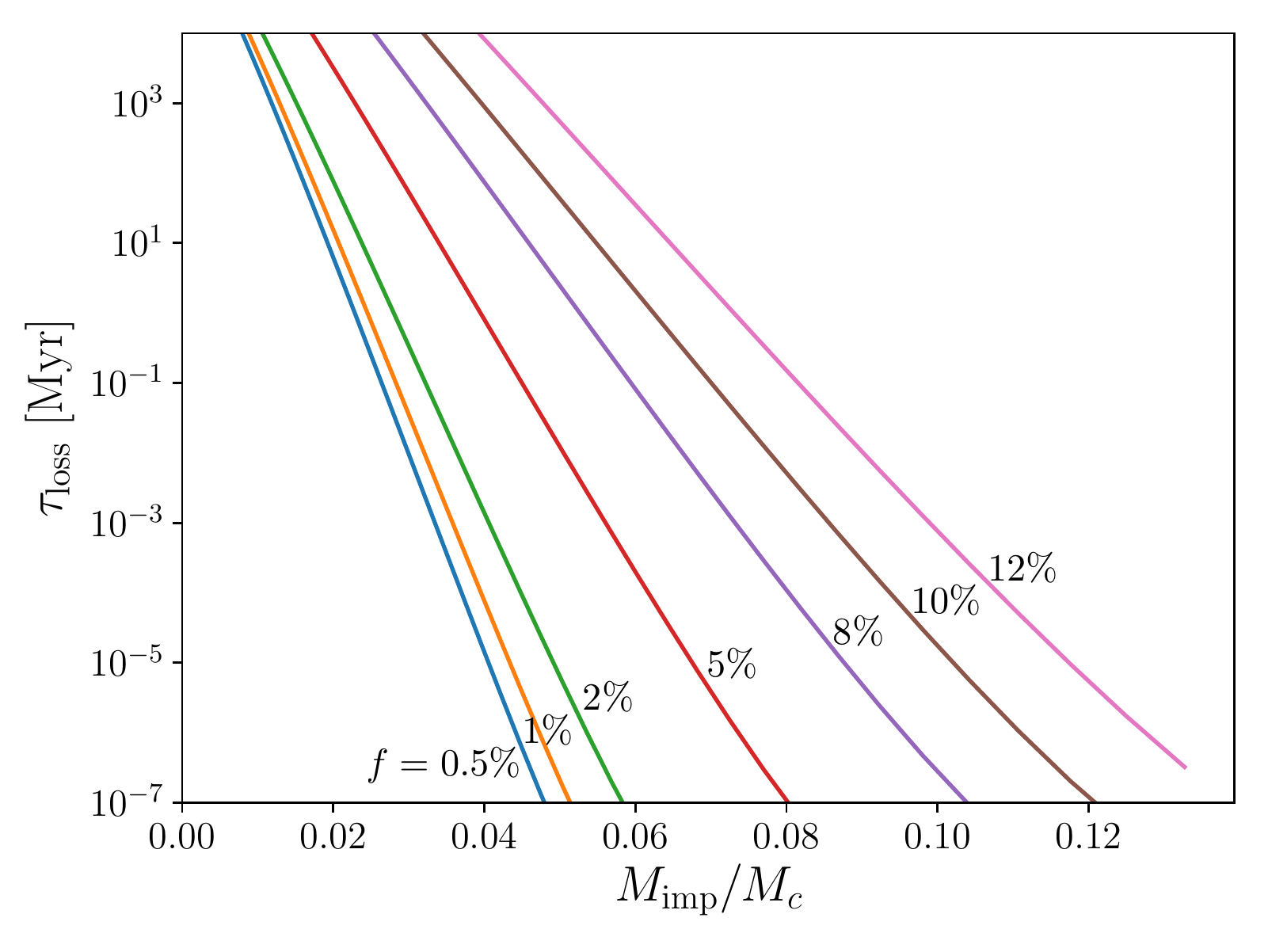}
	\caption{\small Estimated mass loss time-scale ($\tau_\mathrm{loss}$) as a function of impactor mass ($M_\mathrm{imp} / M_c$) for a range of H/He envelope mass fractions ($f$). The initial planet has $M_c = 3 M_\oplus$, $R_c = 1.32 R_\oplus$, $T_{b,0} = 2000~\mathrm{K}$, $a = 0.1~\mathrm{au}$.}
	\label{fig: loss timescale}
\end{figure}

For impacts which do not immediately eject the planetary envelope, significant mass loss is still possible. Taking the mass loss rate (Equation \ref{eq: mdot}) and the exponential density profile in the outer isothermal region (Equation \ref{eq: isothermal density}), the mass loss time-scale, $\tau_\mathrm{loss}$ is
\begin{align}
\label{eq: mass loss timescale}
\tau_\mathrm{loss} & \sim M_\mathrm{env} / \dot{M}_\mathrm{env} \nonumber
\\
& = \frac{M_\mathrm{env}}{4 \pi R_\mathrm{out}^2 c_s \rho_\rcb} \exp{ \left[ \frac{R_\rcb}{h} \left( 1 - \frac{R_\rcb}{R_\mathrm{out}} \right) \right] }
\\
& = \frac{M_\mathrm{env}}{4 \pi R_B^2 c_s \rho_\rcb} \exp{ \left[ \frac{\gamma}{2} \left(\frac{R_B}{R_\rcb} - 1 \right) \right] } \text{,} \nonumber
\end{align}
when $R_\mathrm{out}$ is the Bondi radius, $R_B$, and
\begin{align}
\tau_\mathrm{loss} \sim \frac{M_\mathrm{env}}{4 \pi R_H^2 c_s \rho_\rcb} \exp{ \left[ \frac{\gamma}{2} \left(\frac{R_B}{R_\rcb} - \frac{R_B}{R_H} \right) \right] } \nonumber
\end{align}
when $R_\mathrm{out}$ is the Hill radius, $R_H$. The exponential dependence on $R_\rcb / R_\mathrm{out}$ indicates that the mass loss time-scale is highly sensitive to changes in the location of the radiative-convective boundary relative to either the Bondi or Hill radius.
Therefore, modest changes in the impactor mass, which determines the new atmospheric base temperature, and hence $R_\rcb$, and the semimajor axis, which determines $R_\mathrm{out}$, will strongly affect the mass loss rate (Figure \ref{fig: loss timescale}).

From Equation \eqref{eq: mass loss timescale}, the condition for short mass loss time-scales is $R_\mathrm{out} \sim R_\rcb$, which yields a temperature requirement of
\begin{align}
\label{eq: refined core eject temperature}
T_c = T_\mathrm{eq} \left[ 1 + \left( \frac{R_B}{R_c} - \frac{R_B}{R_\mathrm{out}} \right) \frac{(1 - f) (\gamma - 1)}{2} \right] \text{,}
\end{align}
and an impactor mass of
\begin{align}
\label{eq: refined core eject mass}
M_\mathrm{imp} = \frac{1 - f}{\eta} \nabla_\mathrm{ad} \frac{c_{V,c} \mu}{k_B} \left( \frac{R_c}{R_{\rcb,0}} - \frac{R_c}{R_\mathrm{out}} \right) M_c \text{,}
\end{align}
where the zero subscript indicates the value prior to impact. For planets with $M_c = 3M_\oplus$, $a = 0.1~\mathrm{au}$, and $T_{b,0} = 2000~\mathrm{K}$ (as in Figure \ref{fig: loss timescale}) the impactor must raise the core temperature to ${\sim}10^4~\mathrm{K}$ for rapid loss, corresponding to an impactor mass of ${\sim}0.15 M_\oplus$, or ${\sim}0.05 M_c$.

Short mass loss time-scales are therefore predicted at much lower impactor masses than required for complete loss by Equation \eqref{eq: impactor mass core dominated}. As the envelope sheds mass, it cools, so whether the mass loss can proceed until the envelope is completely lost depends on the relationship of the cooling and mass loss time-scales. For planets with core-dominated energy budgets, the reservoir of energy provided by the core can maintain the envelope in an inflated state, so that rapid mass loss can continue. Since the total energy $E \sim E_c$, the base temperature of the envelope will remain roughly $T_c \sim E_c / (c_{V,c} M_c)$ even as the envelope is lost. In these cases near total atmospheric loss is expected at the impactor masses predicted by Equation \eqref{eq: refined core eject mass}.

\subsection{Results}
To determine the mass loss in a range of impact scenarios and verify our analytical results, we perform numerical integrations of the equations of envelope evolution (Section \ref{sec: envelope evolution}) for a variety of planetary parameters and impactors.
We define the total mass loss fraction, $X$, as the fraction of the envelope's mass lost after $2~\mathrm{Gyr}$ of integration. By $2~\mathrm{Gyr}$ after the impact, contraction of the envelope will have substantially reduced the mass loss rate, so that any additional loss is negligible (Figure \ref{fig: example envelope evolution}). 
Unless otherwise stated, we assume a sunlike star, a core with $M_c = 3M_\oplus$, $R_c / R_\oplus = (M_c / M_\oplus)^{1/4} = 1.32R_\oplus$, and a pre-impact core temperature of $T_{c,0} = 2000~\mathrm{K}$. We assume the H/He envelope has $\mu = 2.34~\mathrm{u}$ and $\gamma = 7/5$, with an atmospheric base temperature that is determined by the core temperature, $T_{b,0} = T_{c,0}$.
We consider a range of semimajor axes and envelope mass fractions, and impactor masses ranging from approximately $0.01{-}0.5 M_\oplus$. We assume that, initially, all the impact energy goes into heating the core, corresponding to an efficiency of $\eta = 1.0$. Additionally, we assume an impact velocity of $v_\mathrm{imp} = v_\mathrm{esc}$. The first assumption is an overestimate, while the second is likely an underestimate. We discuss the implications of these parameter choices in Section \ref{sec: parameter model caveats}.

\begin{figure}
	\centering
	\includegraphics[width=0.5\textwidth]{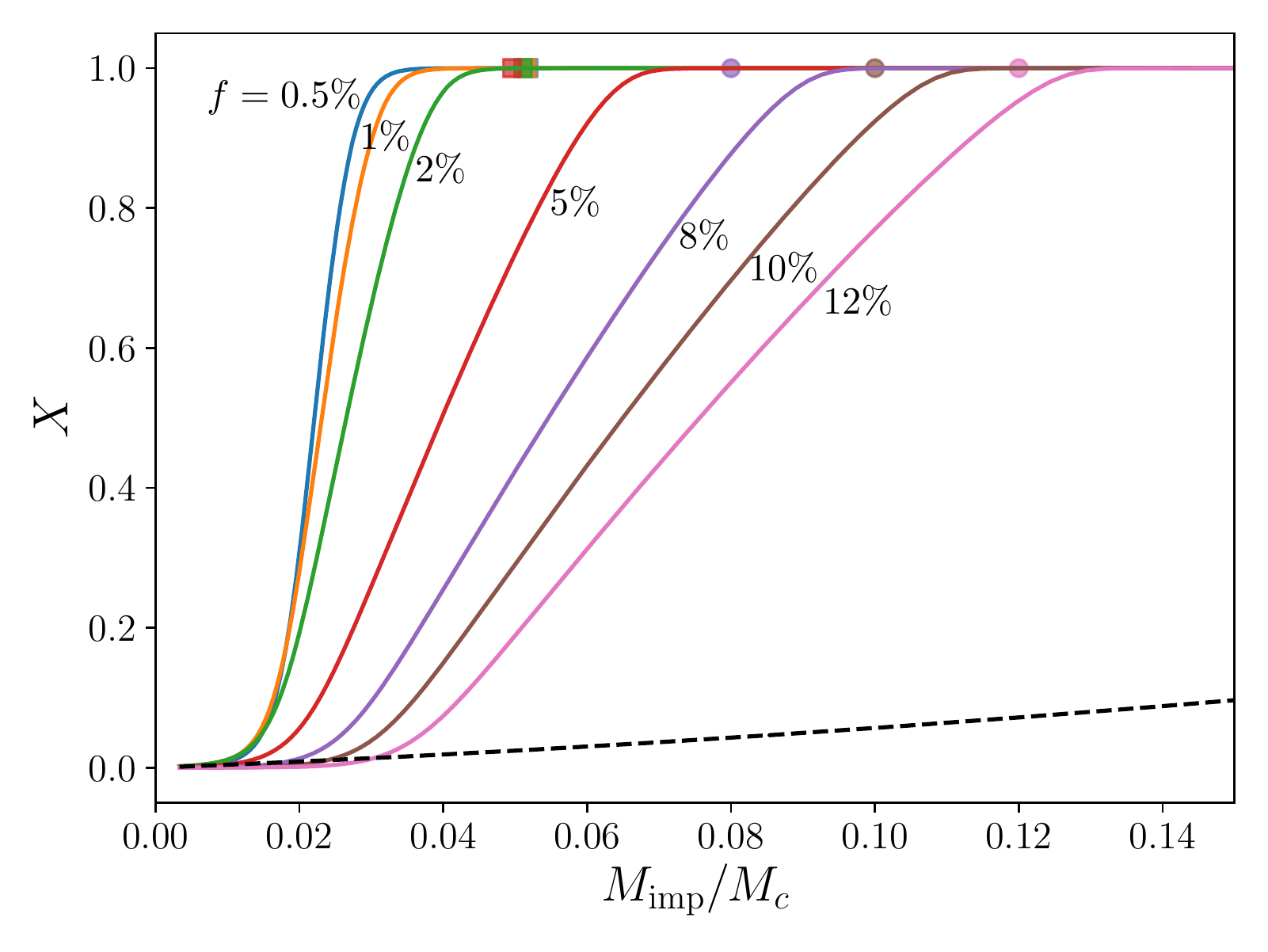}
	\caption{\small Mass fraction of H/He envelope lost ($X$) as a function of impactor mass ($M_\mathrm{imp} / M_c$) for a planet with $M_c = 3 M_\oplus$, $R_c = 1.32 R_\oplus$, $T_{b,0} = 2000~\mathrm{K}$, $a = 0.1~\mathrm{au}$, and a range of envelope mass fractions ($f$). Square and circular marks at $X = 1$ show the analytically derived impactor mass required for total mass loss from Equations \eqref{eq: refined core eject mass} and \eqref{eq: env eject mass} for core- and envelope- dominated energy budgets, respectively. The black dashed line indicates the atmospheric envelope loss from shocks during the impact \citep[Equation (33)]{2015Icar..247...81S} assuming $v_\mathrm{imp} \sim v_\mathrm{esc}$.}
	\label{fig: example loss mass}
\end{figure}

For the reference case, we take $a = 0.1~\mathrm{au}$ and investigate envelope mass fractions ranging from $f = 0.5{-}12$ per cent. The final mass loss for these scenarios is shown in Figure \ref{fig: example loss mass}. We find that atmospheric loss is remarkably efficient. Specifically, depending on $f$, impactor masses of ${\sim}0.1{-}0.4M_\oplus$, or ${\sim}0.03{-}0.13 M_c$, are sufficient to completely remove the envelope. 
For comparison, we include in Figure \ref{fig: example loss mass} the approximate loss caused by hydrodynamic shocks from \citet{2015Icar..247...81S} assuming the same impact velocity  ($v_\mathrm{imp} = v_\mathrm{esc}$).\footnote{The expression here broadly reproduces the results from a more thorough treatment \citep{2015MNRAS.448.1751I, 2016ApJ...817L..13I} and is sufficiently accurate for our purposes.} 
For this size impactor, the hydrodynamic escape caused by thermal expansion is significantly more important than found in previous studies of the shock-induced losses \citep{2016ApJ...817L..13I, 2015MNRAS.448.1751I, 2015Icar..247...81S}.

The higher efficiency of thermal atmospheric loss compared to shock-induced losses can be intuitively understood as follows. For the shock launched by the impactor to completely remove the envelope, the impact has to deliver enough energy and momentum for the ground velocity resulting from the shock as it emerges at the antipode to be comparable to the escape velocity of the planet \citep[for a rough estimate see Equation 28 in][]{2015Icar..247...81S}. For $v_\mathrm{imp} = v_\mathrm{esc}$, this requires impactor masses comparable to the target mass. In contrast, when considering atmospheric losses from the thermal heating of the envelope after the impact, the impact only needs to heat the core to temperatures that yield thermal velocities of the atmosphere sufficient for loss. For H/He atmospheres, this requires less energetic impacts than needed for shock induced volatile loss (Equations \ref{eq: impactor mass core dominated} and \ref{eq: env eject mass}).

\begin{figure}
	\centering
	\includegraphics[width=0.5\textwidth]{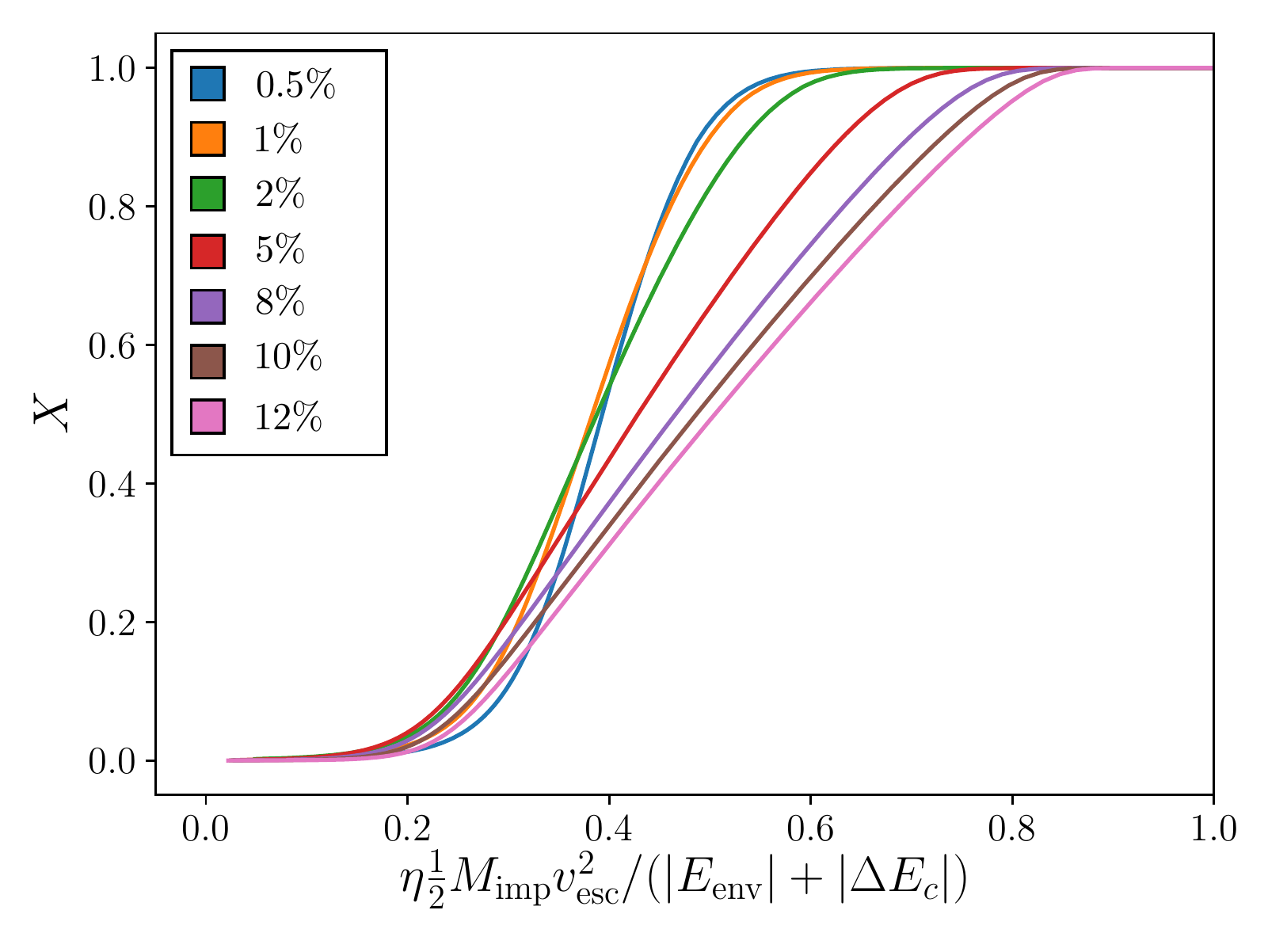}
	\caption{\small Mass fraction of H/He envelope lost ($X$) as a function of normalized impactor energy for a planet with $M_c = 3 M_\oplus$, $R_c = 1.32 R_\oplus$, $T_{b,0} = 2000~\mathrm{K}$, $a = 0.1~\mathrm{au}$, and a range of envelope mass fractions ($f$). The impactor energy is normalized to the sum of the envelope energy ($E_\mathrm{env}$), calculated from Equations \eqref{eq: grav energy} and \eqref{eq: thermal energy}, and the energy required to heat the core to the temperature in Equation \eqref{eq: refined core eject temperature} ($\Delta{E_c}$).}
	\label{fig: example loss energy}
\end{figure}

We find rough agreement between our analytical estimates for the required impactor mass (Equations \ref{eq: env eject mass} and \ref{eq: refined core eject mass}) and our numerical results. In general, our approximate values underestimate the required impactor mass because they are derived for the limiting cases where the energy of the envelope or core can be ignored to first order. 
In reality, the impact energy goes into both the thermal expansion of the envelope and heating the core. 
Including both the finite core and envelope masses by normalizing the impact energy by the envelope energy and core energy deficit, $E_\mathrm{env} + \Delta{E_c}$, we find that the normalized energy required for atmospheric loss is similar across all envelope mass fractions, as shown in Figure \ref{fig: example loss energy}. The envelope energy is calculated from Equations \eqref{eq: grav energy} and \eqref{eq: thermal energy} and $\Delta{E_c}$ is the energy required to heat the core to the temperature in Equation \eqref{eq: refined core eject temperature}.

The analytical expressions overestimate the impactor mass for complete loss in one case: planets on close ($a \lesssim 0.1~\mathrm{au}$) orbits with low-mass envelopes. 
Close-in planets have relatively small Bondi/Hill radii and, because of their higher equilibrium temperatures, larger exponential scale heights ($h \propto t_\mathrm{eq}$). As a consequence, the effect of the exponential decrease in atmospheric density is less severe at the outer radius, and rapid mass loss is possible with lower impactor masses.

\begin{figure}
	\centering
	\includegraphics[width=0.5\textwidth]{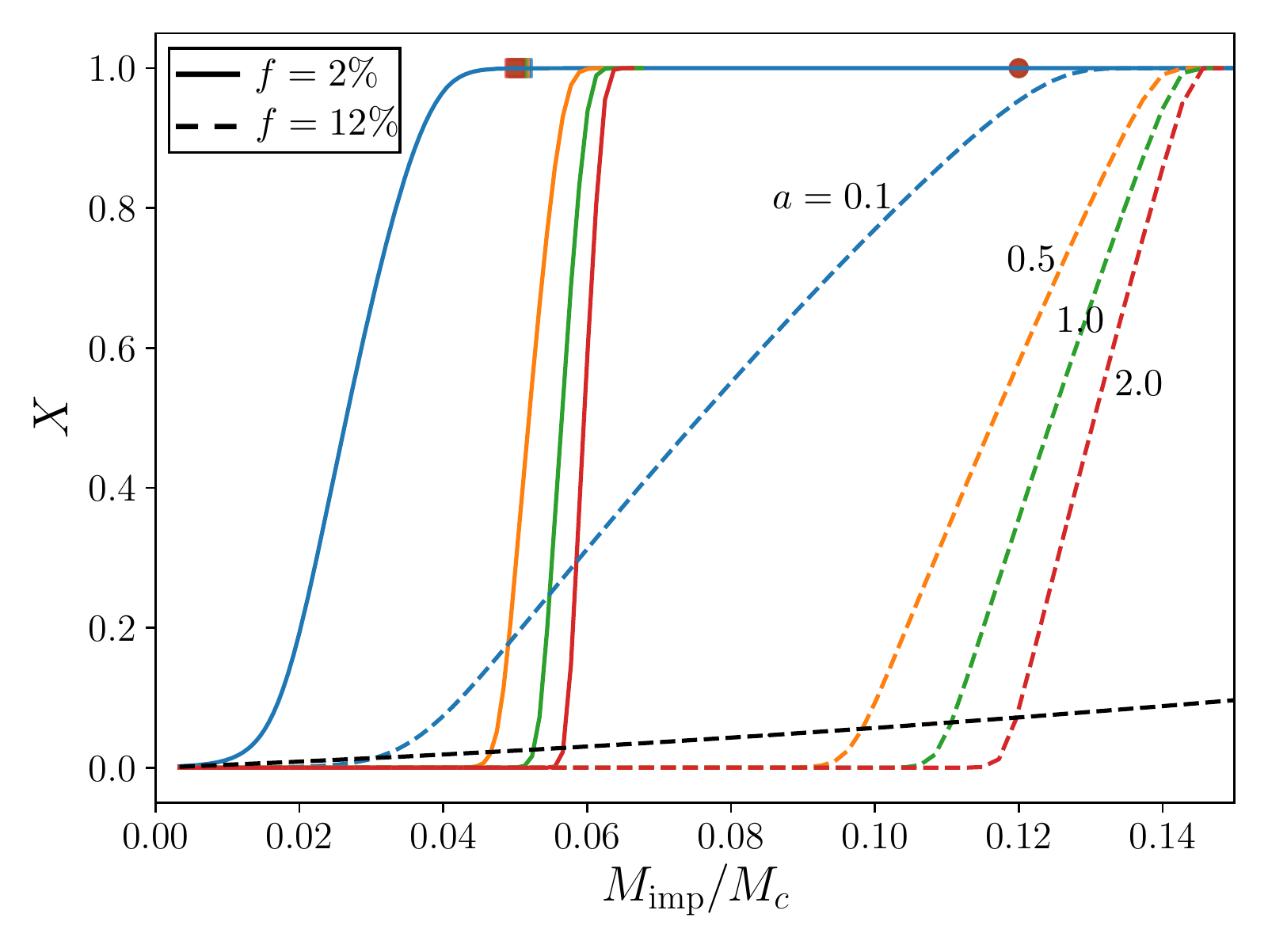}
	\caption{\small Mass fraction of H/He envelope lost ($X$) as a function of impactor mass ($M_\mathrm{imp} / M_c$) for a planet with $M_c = 3 M_\oplus$, $R_c = 1.32 R_\oplus$, $T_{b,0} = 2000~\mathrm{K}$, $f = 2$ or $12$ per cent, and a range of orbital radii ($a$). Square and circular marks at $X = 1$ show the analytically derived impactor mass required for total mass loss from Equations \eqref{eq: refined core eject mass} and \eqref{eq: env eject mass} for core- and envelope- dominated energy budgets, respectively. The black dashed line indicates the atmospheric envelope loss from shocks during the impact \citep[Equation (33)]{2015Icar..247...81S} assuming $v_\mathrm{imp} \sim v_\mathrm{esc}$.}
	\label{fig: loss vs mass with radii}
\end{figure}

\begin{figure}
	\centering
	\includegraphics[width=0.5\textwidth]{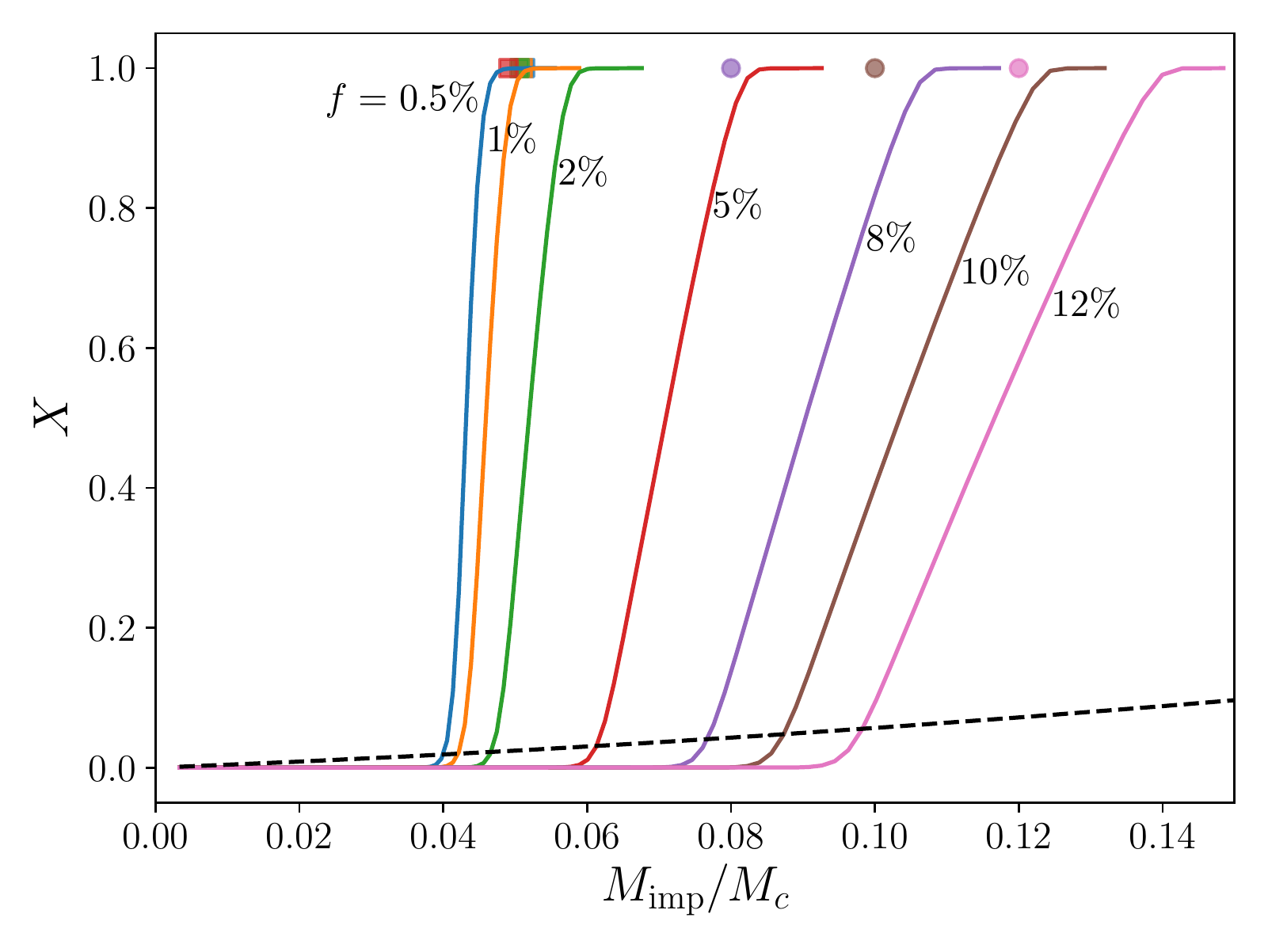}
	\caption{\small Mass fraction of H/He envelope lost ($X$) as a function of impactor mass ($M_\mathrm{imp} / M_c$) for a planet with $M_c = 3 M_\oplus$, $R_c = 1.32 R_\oplus$, $T_{b,0} = 2000~\mathrm{K}$, $a = 0.5~\mathrm{au}$, and a range of envelope mass fractions ($f$). Square and circular marks at $X = 1$ show the analytically derived impactor mass required for total mass loss from Equations \eqref{eq: refined core eject mass} and \eqref{eq: env eject mass} for core- and envelope- dominated energy budgets, respectively. The black dashed line indicates the atmospheric envelope loss from shocks during the impact \citep[Equation (33)]{2015Icar..247...81S} assuming $v_\mathrm{imp} \sim v_\mathrm{esc}$.}
	\label{fig: loss vs mass at 0.5}
\end{figure}

As a planet's orbital separation increases, $R_\mathrm{out}$ ($R_H$ or $R_B$) also increases so that an impact can inflate the envelope without leading to significant mass loss. As shown in Figure \ref{fig: loss vs mass with radii}, this increases the impactor mass required to initiate significant atmospheric loss substantially, but does not dramatically change the impactor mass required for complete loss. As a result, the impactor mass required to initiate loss becomes closer to the mass required for complete ejection, and the range of impactor masses producing intermediate results shrinks, so that most impacts produce binary outcomes--either complete loss or no loss (Figures \ref{fig: loss vs mass with radii} and \ref{fig: loss vs mass at 0.5}).

While this binarity is more pronounced at greater orbital separations, even at $a = 0.1~\mathrm{au}$ the range of impactor masses producing partial envelope loss is relatively narrow. The difference between negligible and total atmospheric loss for a given planet is typically a factor of ${\sim}2{-}4$ in impactor mass, depending on the envelope mass. Regardless of orbital distance, lower mass envelopes have a narrower range of intermediate outcomes and are more likely to entirely lose or retain their envelopes (Figures \ref{fig: example loss mass}, \ref{fig: loss vs mass with radii}, and \ref{fig: loss vs mass at 0.5}).

\begin{figure}
	\centering
	\includegraphics[width=0.5\textwidth]{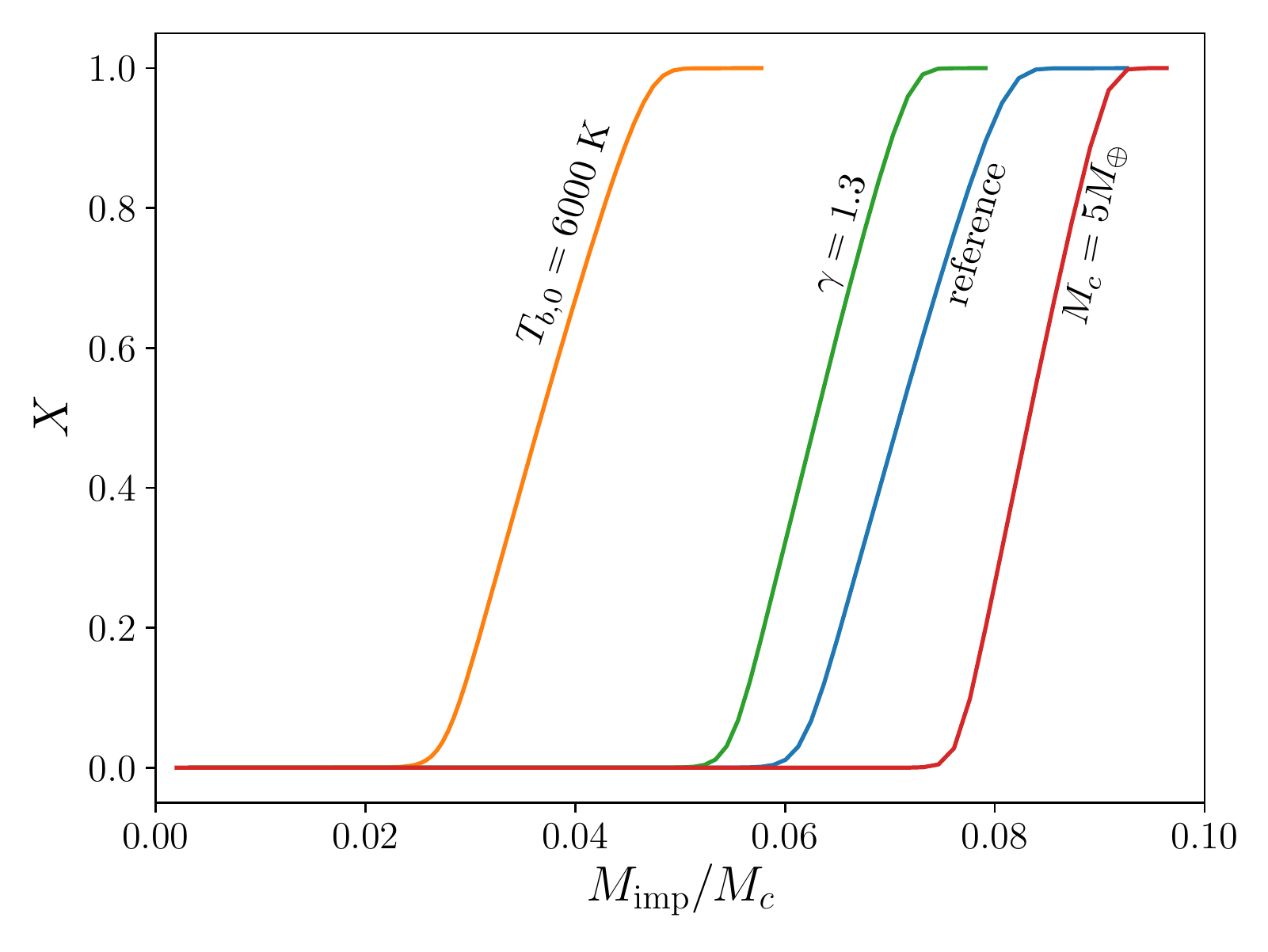}
	\caption{\small Mass fraction of H/He envelope lost ($X$) as a function of impactor mass ($M_\mathrm{imp} / M_c$) under multiple scenarios. The reference scenario is in blue and corresponds to $M_c = 3 M_\oplus$, $R_c = 1.32 R_\oplus$, $a = 0.5~\mathrm{au}$, $f = 5$ per cent, $T_{b,0} = 2000~\mathrm{K}$, and $\gamma=7/5$. In orange, $T_{b,0} \rightarrow 6000~\mathrm{K}$, in green $\gamma \rightarrow 1.3$, and in red $M_c \rightarrow 5 M_\oplus$, while all other parameters match the reference case.}
	\label{fig: loss vs mass for gamma}
\end{figure}

\subsection{Importance of parameter choices and model assumptions}
\label{sec: parameter model caveats}
In the results presented here, we have assumed certain properties for the impact scenario and impacted super-Earth.
We test the effect of changing $\gamma$, $T_{b,0}$, and $M_c$ to determine the sensitivity of our results to our choice of parameters for the target super-Earth (Figure \ref{fig: loss vs mass for gamma}).
Dissociation of molecular hydrogen in the accreting and cooling envelopes of super-Earths and mini-Neptunes can result in values of $\gamma < 4/3$, typically ranging from $1.2 - 1.3$ \citep{2014ApJ...797...95L, 2015ApJ...811...41L, 2016ApJ...817L..13I}. We, therefore, repeated our simulations with a value of $\gamma = 1.3$ instead of $7/5$. This modifies the density profile of the envelope (see Equation \ref{eq: density}) and increases its heat capacity, causing the envelope to receive a larger share of the impact energy. 
The net effect is a marginal improvement in the efficiency of atmospheric removal.

Changes in the base temperature and core mass produce results consistent with expectations from Equations \eqref{eq: env eject mass} and \eqref{eq: refined core eject mass}. Increasing the base temperature of the envelope significantly reduces the impact energy (mass) required to inflate the envelope to the point of atmospheric loss. And, at a fixed $f$, increasing the core mass slightly increases the required impactor-to-core mass ratio ($M_\mathrm{imp} / M_c$) for complete loss. This is because increasing the core mass expands the outer radius while shrinking the portion of the envelope which is convective, resulting in a greater exponential drop in the atmospheric density at the outer radius ($R_B$ or $R_H$) and higher impactor masses to achieve significant atmospheric loss.

We also examine the sensitivity to changes in the impact scenario. In our default scenario, we have conservatively taken $v_\mathrm{imp} = v_\mathrm{esc}$. This is valid, however, only in the case of zero relative motion between the planet and the impactor when they are widely separated. Accounting for this motion yields $v_\mathrm{imp}^2 = v_\mathrm{esc}^2 + v_\infty^2$, where a realistic value for the initial relative motion in the final stages of terrestrial planet formation is $v_\infty \approx v_\mathrm{esc}$ \citep{1976LPSC....7.3245W, 2006Icar..184...39O}.
The result is a doubling of the impact energy expected from a particular event, corresponding to achieving the same atmospheric loss with half the impactor mass.

Conversely, the assumption that $\eta = 1$, i.e. that all of the impact energy is available for heating the core and envelope, yields an upper bound on the thermal atmospheric mass loss. In a real impact, the impact energy is divided into potential, kinetic, and internal energy according the details of the collision. Only the internal energy is available for heating, and only a portion of that internal energy will be converted to heat, with some energy lost to other processes, such as phase transitions. 
Models of a range of Moon-forming impact scenarios suggest that after the shock and decompression phase, $40{-}60$ per cent of the impact energy is left in internal energy \citep{2018LPI....49.2731C}. 
Taking a heating efficiency of $\eta \sim 0.5$ as typical, then the thermal energy of the impact is roughly half what we have assumed above, and the impactor mass required to produce those effects must be doubled.
Impacts which do not result in mergers deposit less energy in the target, resulting in even lower values of $\eta$. These events can be common in environments which are dynamically excited ($v_\infty \gtrsim v_\mathrm{esc}$) \citep[e.g.][]{2010ChEG...70..199A}. But, two bodies which experience such a collision are likely to re-encounter one another, ultimately resulting in an accretionary impact where the bulk of the impact energy is deposited in the target.

The impact velocity and heating efficiency of the impact set the total energy available to power the thermal atmospheric loss, and therefore strongly affect the final outcome. The results presented here, which are equivalent to a choice of $v_\infty \sim v_\mathrm{esc}$ and $\eta \sim 0.5$, are representative of typical outcomes up to a factor of a few. The range of typical outcomes, however, is large and a factor of ${\sim}2{-}4$ change in the thermal energy can make the difference between negligible and complete atmospheric loss (Figure \ref{fig: example loss energy}).

\section{Discussion and Conclusions}
\label{sec: discussion}
We have shown that the thermal expansion of a H/He envelope after a giant impact can lead to a period of rapid mass loss which can substantially erode the atmosphere. We find that, for impactors with a mass ${\sim}10$ per cent of the planet mass, this mechanism can reduce the envelope mass by ${\sim}50{-}100$ per cent. For planets where the thermal energy of the core is much greater than the envelope energy, the impactor mass required for significant atmospheric removal is $M_\mathrm{imp} / M_c \sim \mu / \mu_c \sim 0.1$. This result is only weakly dependent on the atmospheric mass fraction. In contrast, when the envelope energy dominates the total energy budget, the impactor mass required to remove the envelope increases with the envelope mass fraction, $M_\mathrm{imp} \propto M_\mathrm{env}$.
In all cases, however, we find that for H/He envelopes the thermal loss mechanism significantly exceeds the immediate atmospheric loss from impact-generated shocks, which only yields a loss fraction of ${\sim} 10$ per cent for ${\sim}0.1 M_p$ mass impactors \citep[e.g.][]{2016ApJ...817L..13I}.

Our analysis does not include the effect of hydrogen outgassing from the interior of the planet, either directly from the magma ocean or through volcanism, which may act to buffer the atmospheric loss \citep{2018ApJ...854...21C}. We also neglect additional atmospheric mass loss through XUV photoevaporation, which will be enhanced for these planets as a result of their inflated envelopes. The mass loss reported here will, therefore, be an underestimate, though the effect of photoevaporation will be less important for planets with larger semimajor axes \citep{2017ApJ...847...29O}.

The addition of the thermal component of impact-driven atmospheric loss may help to explain the observed diversity in planetary compositions, particularly for planets on orbits outside ${\sim}45~\mathrm{days}$, where photoevaporation and core cooling are less effective, or in systems where neighboring planets display large density contrasts.
Systems of super-Earths are expected to undergo a phase of giant impacts after the dispersal of the gas disc, roughly $10{-}100~\mathrm{Myr}$ after their initial formation \citep{2014A&A...569A..56C, 2017MNRAS.470.1750I}. \citet{2016ApJ...817L..13I} show that the atmospheric loss due to mechanical shocks caused by impacts between similarly-sized bodies can change a planet's bulk density by a factor of $2{-}3$. They suggest this process as an explanation for the observed diversity in planetary bulk density.
We show that the thermal component of the impact should dominate the atmospheric loss of these planets, greatly enhancing the efficiency of atmospheric stripping from impacts. This is especially the case for these impacts because, at an age of $10{-}100~\mathrm{Myr}$, the planetary cores should still retain significant heat from their formation, reducing the impact energy needed to inflate the envelope. Our results suggest that a single large (${\sim}0.1 M_p$) impact is sufficient to remove a H/He atmosphere entirely. Assuming the envelope has a thickness comparable to the core radius, so that $R_p \sim 2 R_c$ \citep[e.g.][]{2014ApJ...792....1L, 2016ApJ...825...29G}, complete atmospheric loss would result in a factor of ${\sim}8$ change in the bulk density.

Atmospheric mass loss is sensitive to the details of the impact scenario (Section \ref{sec: parameter model caveats}). Because these details are stochastic, and because the post-disc instability produces a small number of giant impacts per system \citep{2017MNRAS.470.1750I}, impact-driven mass loss may therefore naturally explain the observed diversity of super-Earth densities.

\section*{Acknowledgements}
This research made use of the software packages NumPy \citep{van2011numpy}, SciPy \citep{jones_scipy_2001}, Astropy \citep{2013A&A...558A..33A}, and matplotlib \citep{Hunter:2007}.
H.E.S gratefully acknowledges support from the National Aeronautics and Space Administration under grant No. 17-XRP17\_2-0055 issued through the Exoplanets Research Program.

\bibliographystyle{mnras}
\bibliography{GiantImpact}

\begin{thebibliography}{}
\makeatletter
\relax
\def\mn@urlcharsother{\let\do\@makeother \do\$\do\&\do\#\do\^\do\_\do\%\do\~}
\def\mn@doi{\begingroup\mn@urlcharsother \@ifnextchar [ {\mn@doi@}
  {\mn@doi@[]}}
\def\mn@doi@[#1]#2{\def\@tempa{#1}\ifx\@tempa\@empty \href
  {http://dx.doi.org/#2} {doi:#2}\else \href {http://dx.doi.org/#2} {#1}\fi
  \endgroup}
\def\mn@eprint#1#2{\mn@eprint@#1:#2::\@nil}
\def\mn@eprint@arXiv#1{\href {http://arxiv.org/abs/#1} {{\tt arXiv:#1}}}
\def\mn@eprint@dblp#1{\href {http://dblp.uni-trier.de/rec/bibtex/#1.xml}
  {dblp:#1}}
\def\mn@eprint@#1:#2:#3:#4\@nil{\def\@tempa {#1}\def\@tempb {#2}\def\@tempc
  {#3}\ifx \@tempc \@empty \let \@tempc \@tempb \let \@tempb \@tempa \fi \ifx
  \@tempb \@empty \def\@tempb {arXiv}\fi \@ifundefined
  {mn@eprint@\@tempb}{\@tempb:\@tempc}{\expandafter \expandafter \csname
  mn@eprint@\@tempb\endcsname \expandafter{\@tempc}}}

\bibitem[\protect\citeauthoryear{{Adams}, {Seager}  \& {Elkins-Tanton}}{{Adams}
  et~al.}{2008}]{2008ApJ...673.1160A}
{Adams} E.~R.,  {Seager} S.,   {Elkins-Tanton} L.,  2008, \mn@doi [\apj]
  {10.1086/524925}, \href {http://adsabs.harvard.edu/abs/2008ApJ...673.1160A}
  {673, 1160}

\bibitem[\protect\citeauthoryear{{Alf{\`e}}, {Price}  \& {Gillan}}{{Alf{\`e}}
  et~al.}{2001}]{2001PhRvB..64d5123A}
{Alf{\`e}} D.,  {Price} G.~D.,   {Gillan} M.~J.,  2001, \mn@doi [\prb]
  {10.1103/PhysRevB.64.045123}, \href
  {http://adsabs.harvard.edu/abs/2001PhRvB..64d5123A} {64, 045123}

\bibitem[\protect\citeauthoryear{{Asphaug}}{{Asphaug}}{2010}]{2010ChEG...70..199A}
{Asphaug} E.,  2010, \mn@doi [Chemie der Erde / Geochemistry]
  {10.1016/j.chemer.2010.01.004}, \href
  {https://ui.adsabs.harvard.edu/#abs/2010ChEG...70..199A} {70, 199}

\bibitem[\protect\citeauthoryear{{Astropy Collaboration} et~al.,}{{Astropy
  Collaboration} et~al.}{2013}]{2013A&A...558A..33A}
{Astropy Collaboration} et~al., 2013, \mn@doi [\aap]
  {10.1051/0004-6361/201322068}, \href
  {http://adsabs.harvard.edu/abs/2013A%26A...558A..33A} {558, A33}

\bibitem[\protect\citeauthoryear{{Borucki} et~al.,}{{Borucki}
  et~al.}{2011}]{2011ApJ...736...19B}
{Borucki} W.~J.,  et~al., 2011, \mn@doi [\apj] {10.1088/0004-637X/736/1/19},
  \href {http://adsabs.harvard.edu/abs/2011ApJ...736...19B} {736, 19}

\bibitem[\protect\citeauthoryear{{Carter}, {Lock}  \& {Stewart}}{{Carter}
  et~al.}{2018}]{2018LPI....49.2731C}
{Carter} P.~J.,  {Lock} S.~J.,   {Stewart} S.~T.,  2018, in Lunar and Planetary
  Science Conference. p.~2731

\bibitem[\protect\citeauthoryear{{Chachan} \& {Stevenson}}{{Chachan} \&
  {Stevenson}}{2018}]{2018ApJ...854...21C}
{Chachan} Y.,  {Stevenson} D.~J.,  2018, \mn@doi [\apj]
  {10.3847/1538-4357/aaa459}, \href
  {https://ui.adsabs.harvard.edu/#abs/2018ApJ...854...21C} {854, 21}

\bibitem[\protect\citeauthoryear{{Chambers}}{{Chambers}}{2001}]{2001Icar..152..205C}
{Chambers} J.~E.,  2001, \mn@doi [\icarus] {10.1006/icar.2001.6639}, \href
  {https://ui.adsabs.harvard.edu/#abs/2001Icar..152..205C} {152, 205}

\bibitem[\protect\citeauthoryear{{Cossou}, {Raymond}, {Hersant}  \&
  {Pierens}}{{Cossou} et~al.}{2014}]{2014A&A...569A..56C}
{Cossou} C.,  {Raymond} S.~N.,  {Hersant} F.,   {Pierens} A.,  2014, \mn@doi
  [\aap] {10.1051/0004-6361/201424157}, \href
  {http://adsabs.harvard.edu/abs/2014A%26A...569A..56C} {569, A56}

\bibitem[\protect\citeauthoryear{{Denham}, {Naoz}, {Hoang}, {Stephan}  \&
  {Farr}}{{Denham} et~al.}{2018}]{2018arXiv180200447D}
{Denham} P.,  {Naoz} S.,  {Hoang} B.-M.,  {Stephan} A.~P.,   {Farr} W.~M.,
  2018, preprint, \href {http://adsabs.harvard.edu/abs/2018arXiv180200447D} {}
  (\mn@eprint {arXiv} {1802.00447})

\bibitem[\protect\citeauthoryear{{Freedman}, {Marley}  \& {Lodders}}{{Freedman}
  et~al.}{2008}]{2008ApJS..174..504F}
{Freedman} R.~S.,  {Marley} M.~S.,   {Lodders} K.,  2008, \mn@doi [\apjs]
  {10.1086/521793}, \href {http://adsabs.harvard.edu/abs/2008ApJS..174..504F}
  {174, 504}

\bibitem[\protect\citeauthoryear{{Fressin} et~al.,}{{Fressin}
  et~al.}{2013}]{2013ApJ...766...81F}
{Fressin} F.,  et~al., 2013, \mn@doi [\apj] {10.1088/0004-637X/766/2/81}, \href
  {http://adsabs.harvard.edu/abs/2013ApJ...766...81F} {766, 81}

\bibitem[\protect\citeauthoryear{{Fulton} \& {Petigura}}{{Fulton} \&
  {Petigura}}{2018}]{2018arXiv180501453F}
{Fulton} B.~J.,  {Petigura} E.~A.,  2018, preprint, \href
  {https://ui.adsabs.harvard.edu/#abs/2018arXiv180501453F} {p.
  arXiv:1805.01453} (\mn@eprint {arXiv} {1805.01453})

\bibitem[\protect\citeauthoryear{{Fulton} et~al.,}{{Fulton}
  et~al.}{2017}]{2017AJ....154..109F}
{Fulton} B.~J.,  et~al., 2017, \mn@doi [\aj] {10.3847/1538-3881/aa80eb}, \href
  {http://adsabs.harvard.edu/abs/2017AJ....154..109F} {154, 109}

\bibitem[\protect\citeauthoryear{{Genda} \& {Abe}}{{Genda} \&
  {Abe}}{2003}]{2003Icar..164..149G}
{Genda} H.,  {Abe} Y.,  2003, \mn@doi [\icarus]
  {10.1016/S0019-1035(03)00101-5}, \href
  {http://adsabs.harvard.edu/abs/2003Icar..164..149G} {164, 149}

\bibitem[\protect\citeauthoryear{{Ginzburg}, {Schlichting}  \&
  {Sari}}{{Ginzburg} et~al.}{2016}]{2016ApJ...825...29G}
{Ginzburg} S.,  {Schlichting} H.~E.,   {Sari} R.,  2016, \mn@doi [\apj]
  {10.3847/0004-637X/825/1/29}, \href
  {http://adsabs.harvard.edu/abs/2016ApJ...825...29G} {825, 29}

\bibitem[\protect\citeauthoryear{{Ginzburg}, {Schlichting}  \&
  {Sari}}{{Ginzburg} et~al.}{2018}]{2018MNRAS.476..759G}
{Ginzburg} S.,  {Schlichting} H.~E.,   {Sari} R.,  2018, \mn@doi [\mnras]
  {10.1093/mnras/sty290}, \href
  {http://adsabs.harvard.edu/abs/2018MNRAS.476..759G} {476, 759}

\bibitem[\protect\citeauthoryear{{Guillot}, {Chabrier}, {Gautier}  \&
  {Morel}}{{Guillot} et~al.}{1995}]{1995ApJ...450..463G}
{Guillot} T.,  {Chabrier} G.,  {Gautier} D.,   {Morel} P.,  1995, \mn@doi
  [\apj] {10.1086/176156}, \href
  {http://adsabs.harvard.edu/abs/1995ApJ...450..463G} {450, 463}

\bibitem[\protect\citeauthoryear{{Hansen} \& {Murray}}{{Hansen} \&
  {Murray}}{2013}]{2013ApJ...775...53H}
{Hansen} B.~M.~S.,  {Murray} N.,  2013, \mn@doi [\apj]
  {10.1088/0004-637X/775/1/53}, \href
  {http://adsabs.harvard.edu/abs/2013ApJ...775...53H} {775, 53}

\bibitem[\protect\citeauthoryear{Hunter}{Hunter}{2007}]{Hunter:2007}
Hunter J.~D.,  2007, Computing In Science \& Engineering, 9, 90

\bibitem[\protect\citeauthoryear{{Inamdar} \& {Schlichting}}{{Inamdar} \&
  {Schlichting}}{2015}]{2015MNRAS.448.1751I}
{Inamdar} N.~K.,  {Schlichting} H.~E.,  2015, \mn@doi [\mnras]
  {10.1093/mnras/stv030}, \href
  {http://adsabs.harvard.edu/abs/2015MNRAS.448.1751I} {448, 1751}

\bibitem[\protect\citeauthoryear{{Inamdar} \& {Schlichting}}{{Inamdar} \&
  {Schlichting}}{2016}]{2016ApJ...817L..13I}
{Inamdar} N.~K.,  {Schlichting} H.~E.,  2016, \mn@doi [\apjl]
  {10.3847/2041-8205/817/2/L13}, \href
  {http://adsabs.harvard.edu/abs/2016ApJ...817L..13I} {817, L13}

\bibitem[\protect\citeauthoryear{{Izidoro}, {Ogihara}, {Raymond}, {Morbidelli},
  {Pierens}, {Bitsch}, {Cossou}  \& {Hersant}}{{Izidoro}
  et~al.}{2017}]{2017MNRAS.470.1750I}
{Izidoro} A.,  {Ogihara} M.,  {Raymond} S.~N.,  {Morbidelli} A.,  {Pierens} A.,
   {Bitsch} B.,  {Cossou} C.,   {Hersant} F.,  2017, \mn@doi [\mnras]
  {10.1093/mnras/stx1232}, \href
  {http://adsabs.harvard.edu/abs/2017MNRAS.470.1750I} {470, 1750}

\bibitem[\protect\citeauthoryear{{Jin}, {Mordasini}, {Parmentier}, {van
  Boekel}, {Henning}  \& {Ji}}{{Jin} et~al.}{2014}]{2014ApJ...795...65J}
{Jin} S.,  {Mordasini} C.,  {Parmentier} V.,  {van Boekel} R.,  {Henning} T.,
  {Ji} J.,  2014, \mn@doi [\apj] {10.1088/0004-637X/795/1/65}, \href
  {http://adsabs.harvard.edu/abs/2014ApJ...795...65J} {795, 65}

\bibitem[\protect\citeauthoryear{{Katsura} et~al.,}{{Katsura}
  et~al.}{2010}]{2010PEPI..183..212K}
{Katsura} T.,  et~al., 2010, \mn@doi [Physics of the Earth and Planetary
  Interiors] {10.1016/j.pepi.2010.07.001}, \href
  {http://adsabs.harvard.edu/abs/2010PEPI..183..212K} {183, 212}

\bibitem[\protect\citeauthoryear{{Lee} \& {Chiang}}{{Lee} \&
  {Chiang}}{2015}]{2015ApJ...811...41L}
{Lee} E.~J.,  {Chiang} E.,  2015, \mn@doi [\apj] {10.1088/0004-637X/811/1/41},
  \href {http://adsabs.harvard.edu/abs/2015ApJ...811...41L} {811, 41}

\bibitem[\protect\citeauthoryear{{Lee}, {Chiang}  \& {Ormel}}{{Lee}
  et~al.}{2014}]{2014ApJ...797...95L}
{Lee} E.~J.,  {Chiang} E.,   {Ormel} C.~W.,  2014, \mn@doi [\apj]
  {10.1088/0004-637X/797/2/95}, \href
  {http://adsabs.harvard.edu/abs/2014ApJ...797...95L} {797, 95}

\bibitem[\protect\citeauthoryear{{Liu}, {Hori}, {Lin}  \& {Asphaug}}{{Liu}
  et~al.}{2015}]{2015ApJ...812..164L}
{Liu} S.-F.,  {Hori} Y.,  {Lin} D.~N.~C.,   {Asphaug} E.,  2015, \mn@doi [\apj]
  {10.1088/0004-637X/812/2/164}, \href
  {http://adsabs.harvard.edu/abs/2015ApJ...812..164L} {812, 164}

\bibitem[\protect\citeauthoryear{{Lopez} \& {Fortney}}{{Lopez} \&
  {Fortney}}{2013}]{2013ApJ...776....2L}
{Lopez} E.~D.,  {Fortney} J.~J.,  2013, \mn@doi [\apj]
  {10.1088/0004-637X/776/1/2}, \href
  {http://adsabs.harvard.edu/abs/2013ApJ...776....2L} {776, 2}

\bibitem[\protect\citeauthoryear{{Lopez} \& {Fortney}}{{Lopez} \&
  {Fortney}}{2014}]{2014ApJ...792....1L}
{Lopez} E.~D.,  {Fortney} J.~J.,  2014, \mn@doi [\apj]
  {10.1088/0004-637X/792/1/1}, \href
  {http://adsabs.harvard.edu/abs/2014ApJ...792....1L} {792, 1}

\bibitem[\protect\citeauthoryear{{Lopez}, {Fortney}  \& {Miller}}{{Lopez}
  et~al.}{2012}]{2012ApJ...761...59L}
{Lopez} E.~D.,  {Fortney} J.~J.,   {Miller} N.,  2012, \mn@doi [\apj]
  {10.1088/0004-637X/761/1/59}, \href
  {http://adsabs.harvard.edu/abs/2012ApJ...761...59L} {761, 59}

\bibitem[\protect\citeauthoryear{{Morton}, {Bryson}, {Coughlin}, {Rowe},
  {Ravichandran}, {Petigura}, {Haas}  \& {Batalha}}{{Morton}
  et~al.}{2016}]{2016ApJ...822...86M}
{Morton} T.~D.,  {Bryson} S.~T.,  {Coughlin} J.~L.,  {Rowe} J.~F.,
  {Ravichandran} G.,  {Petigura} E.~A.,  {Haas} M.~R.,   {Batalha} N.~M.,
  2016, \mn@doi [\apj] {10.3847/0004-637X/822/2/86}, \href
  {http://adsabs.harvard.edu/abs/2016ApJ...822...86M} {822, 86}

\bibitem[\protect\citeauthoryear{{O'Brien}, {Morbidelli}  \&
  {Levison}}{{O'Brien} et~al.}{2006}]{2006Icar..184...39O}
{O'Brien} D.~P.,  {Morbidelli} A.,   {Levison} H.~F.,  2006, \mn@doi [\icarus]
  {10.1016/j.icarus.2006.04.005}, \href
  {http://adsabs.harvard.edu/abs/2006Icar..184...39O} {184, 39}

\bibitem[\protect\citeauthoryear{{Owen} \& {Wu}}{{Owen} \&
  {Wu}}{2013}]{2013ApJ...775..105O}
{Owen} J.~E.,  {Wu} Y.,  2013, \mn@doi [\apj] {10.1088/0004-637X/775/2/105},
  \href {http://adsabs.harvard.edu/abs/2013ApJ...775..105O} {775, 105}

\bibitem[\protect\citeauthoryear{{Owen} \& {Wu}}{{Owen} \&
  {Wu}}{2016}]{2016ApJ...817..107O}
{Owen} J.~E.,  {Wu} Y.,  2016, \mn@doi [\apj] {10.3847/0004-637X/817/2/107},
  \href {http://adsabs.harvard.edu/abs/2016ApJ...817..107O} {817, 107}

\bibitem[\protect\citeauthoryear{{Owen} \& {Wu}}{{Owen} \&
  {Wu}}{2017}]{2017ApJ...847...29O}
{Owen} J.~E.,  {Wu} Y.,  2017, \mn@doi [\apj] {10.3847/1538-4357/aa890a}, \href
  {http://adsabs.harvard.edu/abs/2017ApJ...847...29O} {847, 29}

\bibitem[\protect\citeauthoryear{{Petigura}, {Howard}  \& {Marcy}}{{Petigura}
  et~al.}{2013}]{2013PNAS..11019273P}
{Petigura} E.~A.,  {Howard} A.~W.,   {Marcy} G.~W.,  2013, \mn@doi [Proceedings
  of the National Academy of Science] {10.1073/pnas.1319909110}, \href
  {http://adsabs.harvard.edu/abs/2013PNAS..11019273P} {110, 19273}

\bibitem[\protect\citeauthoryear{{Piso} \& {Youdin}}{{Piso} \&
  {Youdin}}{2014}]{2014ApJ...786...21P}
{Piso} A.-M.~A.,  {Youdin} A.~N.,  2014, \mn@doi [\apj]
  {10.1088/0004-637X/786/1/21}, \href
  {http://adsabs.harvard.edu/abs/2014ApJ...786...21P} {786, 21}

\bibitem[\protect\citeauthoryear{{Rafikov}}{{Rafikov}}{2006}]{2006ApJ...648..666R}
{Rafikov} R.~R.,  2006, \mn@doi [\apj] {10.1086/505695}, \href
  {http://adsabs.harvard.edu/abs/2006ApJ...648..666R} {648, 666}

\bibitem[\protect\citeauthoryear{{Rogers}}{{Rogers}}{2015}]{2015ApJ...801...41R}
{Rogers} L.~A.,  2015, \mn@doi [\apj] {10.1088/0004-637X/801/1/41}, \href
  {http://adsabs.harvard.edu/abs/2015ApJ...801...41R} {801, 41}

\bibitem[\protect\citeauthoryear{{Schlichting}}{{Schlichting}}{2018}]{2018haex.bookE.141S}
{Schlichting} H.,  2018, {Formation of Super-Earths}.
p.~141, \mn@doi{10.1007/978-3-319-30648-3_141-1}

\bibitem[\protect\citeauthoryear{{Schlichting}, {Sari}  \&
  {Yalinewich}}{{Schlichting} et~al.}{2015}]{2015Icar..247...81S}
{Schlichting} H.~E.,  {Sari} R.,   {Yalinewich} A.,  2015, \mn@doi [\icarus]
  {10.1016/j.icarus.2014.09.053}, \href
  {http://adsabs.harvard.edu/abs/2015Icar..247...81S} {247, 81}

\bibitem[\protect\citeauthoryear{{Valencia}, {O'Connell}  \&
  {Sasselov}}{{Valencia} et~al.}{2006}]{2006Icar..181..545V}
{Valencia} D.,  {O'Connell} R.~J.,   {Sasselov} D.,  2006, \mn@doi [\icarus]
  {10.1016/j.icarus.2005.11.021}, \href
  {https://ui.adsabs.harvard.edu/#abs/2006Icar..181..545V} {181, 545}

\bibitem[\protect\citeauthoryear{Van Der~Walt, Colbert  \& Varoquaux}{Van
  Der~Walt et~al.}{2011}]{van2011numpy}
Van Der~Walt S.,  Colbert S.~C.,   Varoquaux G.,  2011, Computing in Science \&
  Engineering, 13, 22

\bibitem[\protect\citeauthoryear{{Weiss} \& {Marcy}}{{Weiss} \&
  {Marcy}}{2014}]{2014ApJ...783L...6W}
{Weiss} L.~M.,  {Marcy} G.~W.,  2014, \mn@doi [\apjl]
  {10.1088/2041-8205/783/1/L6}, \href
  {http://adsabs.harvard.edu/abs/2014ApJ...783L...6W} {783, L6}

\bibitem[\protect\citeauthoryear{{Wetherill}}{{Wetherill}}{1976}]{1976LPSC....7.3245W}
{Wetherill} G.~W.,  1976, in {Merrill} R.~B.,  ed.,  Lunar and Planetary
  Science Conference Proceedings Vol. 7, Lunar and Planetary Science Conference
  Proceedings. pp 3245--3257

\bibitem[\protect\citeauthoryear{Van Der~Walt, Colbert  \&
  Varoquaux}{jon}{2001}]{jones_scipy_2001}
 2001, {SciPy}: Open source scientific tools for Python, \url
  {http://www.scipy.org/}

\makeatother
\end{thebibliography}
\bsp
\label{lastpage}
\end{document}